\documentclass[12pt]{article}

\usepackage{latexsym} 
\usepackage{amssymb}  
\usepackage{amsbsy}   
\usepackage{graphicx}       
\usepackage[dvips]{color}
\usepackage{bm}   

\newcommand{\al}{\alpha}
\newcommand{\be}{\beta}
\newcommand{\ga}{\gamma}

\newcommand{\de}{\delta}
\newcommand{\De}{\Delta}

\newcommand{\eps}{\epsilon}

\newcommand{\ka}{\kappa}

\newcommand{\La}{\Lambda}

\newcommand{\si}{\sigma}

\newcommand{\beq}{\begin{equation}}
\newcommand{\eeq}{\end{equation}}
\newcommand{\ba}{\begin{array}}
\newcommand{\ea}{\end{array}}
\newcommand{\bea}{\begin{eqnarray}}
\newcommand{\eea}{\end{eqnarray}}
\newcommand{\bi}{\begin{itemize}}  
\newcommand{\ei}{\end{itemize}}
\newcommand{\ben}{\begin{enumerate}} 
\newcommand{\een}{\end{enumerate}}
\newcommand{\bc}{\begin{center}}
\newcommand{\ec}{\end{center}}

\newcommand{\p}{\partial}
 
\newcommand{\txt}{\textstyle}

\newcommand\eqn[1]{(\ref{#1})}      

\newcommand{\half} {{\txt \frac{1}{2}}}

\newcommand{\ee}[1]{\times 10^{#1}}

\newcommand{\MeV}{{\rm MeV}}

\newcommand{\one}{\bm{1}}
\newcommand{\dmu}{\delta\mu}
\newcommand{\mubar}{{\bar\mu}}

\usepackage[normalem]{ulem}  
\newcommand{\new}[1]{#1}

\makeatletter 

\def\appendix{\par                              
    \setcounter{section}{0}                     
    \setcounter{subsection}{0}
    \renewcommand{\theequation}{\Alph{section}.\arabic{equation}}
    \renewcommand{\thesection}{Appendix \Alph{section}
                \setcounter{equation}{0}  } 
    \renewcommand{\thesubsection}{\Alph{section}.\arabic{subsection}}
}

\def\applabel#1{\@bsphack
  \protected@write\@auxout{}%
         {\string\newlabel{#1}{{\Alph{section}}{\thepage}}}%
  \@esphack}

\def\section{
\setcounter{equation}{0}        
\@startsection {section}{1}{\z@}{-3.5ex plus -1ex minus 
 -.2ex}{2.3ex plus .2ex}{\large\bf}}
\renewcommand{\theequation}{\arabic{section}.\arabic{equation}}

\def\subsection{\@startsection{subsection}{2}{\z@}{-3.25ex plus -1ex minus 
 -.2ex}{1.5ex plus .2ex}{\normalsize\bf}}

\def\subsubsection{\@startsection{subsubsection}{3}{\z@}{-3.25ex plus
 -1ex minus -.2ex}{1.5ex plus .2ex}{\normalsize}}

\makeatother   

\begin{document}

 
\title{\bf Secondary pairing in gapless color-superconducting quark matter}

\author{
Mark Alford and Qing-hai Wang \\[2ex]
Department of Physics \\ Washington University \\
St.~Louis, MO~63130 \\ USA}

\date{July 22, 2005}

\begin{titlepage}
\maketitle
\renewcommand{\thepage}{}          

\begin{abstract}
We calculate secondary pairing in a model of a color superconductor
with a quadratic gapless dispersion relation for the quasiquarks of
the primary pairing.  Our model mimics the physics of the sector of
blue up and red strange quarks in gapless color-flavor-locked quark
matter.  The secondary pairing opens up a gap $\Delta_s$ in the 
quark spectrum, 
and we confirm Hong's prediction that in typical secondary
channels $\Delta_s \propto G_s^2$ for coupling strength $G_s$.  This
shows that the large density of states of the quadratically gapless mode
greatly enhances the secondary pairing over the standard BCS result
$\Delta \propto \exp(-{\rm const}/G)$.  
In all of the secondary channels that we
analyzed we find that the secondary gap, even with this enhancement,
is from ten to hundreds of
times smaller than the primary gap at reasonable values of the
secondary coupling, indicating that secondary pairing
does not generically resolve the magnetic
instability of the gapless phase.
\end{abstract}

\end{titlepage}


\section{Introduction}
\label{sec:intro}

The exploration of the phase diagram of matter at ultra-high temperature
or density is an area of great interest and activity, both on the
experimental and theoretical fronts. Heavy-ion colliders such as
the SPS at CERN and RHIC at Brookhaven have probed the high-temperature
region, searching for the transition to deconfined quark matter.
In this paper we discuss a puzzle that arises in a different part of
the phase diagram, at low temperature but ultra-high density. Here there are
as yet no experimental constraints, but calculations
show that at sufficiently
high density, the favored phase is color-flavor-locked (CFL)
color-superconducting quark matter \cite{CFL} (for reviews,
see Ref.~\cite{Reviews}). 

The puzzle concerns the identity of
the next phase down in density. Recent
work \cite{gCFL} suggests that when the density drops low enough so that
the mass of the strange quark can no longer be neglected, there
is a continuous phase transition from the CFL phase to a new
gapless CFL (gCFL) phase, which could lead to observable consequences
if it occurred in the cores of neutron stars \cite{Alford:2004zr}.
However, it now appears that some of the gluons
in the gCFL phase have imaginary
Meissner masses, indicating an instability towards 
an unknown lower-energy phase
\cite{Huang:2004am,Casalbuoni:2004tb,Giannakis:2004pf,Fukushima:2005cm}. 

The gCFL phase is named after its most striking characteristic:
the presence of gapless modes in the spectrum of quark excitations
above the color-superconducting ground state. These include
a gapless mode
with an approximately quadratic dispersion relation $E(p) \propto (p-p_F)^2$,
as well as gapless modes with the more typical linear dispersion relation
$E(p) \propto |p-p_F|$. 
It seems likely that the presence of these modes is related to the
instability\footnote{Note, however, that the instability can occur
in fully gapped superconductors \cite{Huang:2004am},
for example at finite temperature \cite{Alford:2005qw}.}.
It has therefore been suggested that one way to resolve the
instability would be for the gapless quasiparticles to pair with each other
(``secondary pairing''), leaving no gapless modes
at all. In particular, it has been argued \cite{Hong} that
because the quadratic gapless mode has so much phase space at low energy 
(its density of states diverges as $E^{-1/2}$) the secondary pairing will be
greatly enhanced over the standard BCS value, which is based on pairing
of modes that are linearly gapless. This offers the prospect that channels 
whose attraction seemed negligibly weak could become important once the
primary pairing has created a quadratically gapless 
quasiquark, and these channels could then supply
the required secondary pairing. It must be remembered, 
however, that in order to fully resolve the instability the secondary pairing
gap parameter $\De_s$ must be comparable to the primary pairing $\De_p$.
If $\De_s$ were significantly smaller then there would be
a temperature range $\De_s \ll T \ll \De_p$ in which there was primary
pairing but no secondary pairing, and at those temperatures the
instability problem would arise again.

In this paper we investigate secondary pairing in a simple two-species
model with a Nambu--Jona-Lasinio (NJL) interaction. We previously
used a similar model to study the effects of gapless modes on
photon and gluon screening masses \cite{Alford:2005qw}.
Our model is essentially just one sector of the gCFL pairing
pattern (the blue-up/red-strange sector), which is where
a quadratically gapless quasiquark emerges, after
primary pairing between the two species, in the Dirac $C\ga_5$ channel.

The Cooper pairs must have an overall antisymmetric wavefunction,
and because the secondary pairing is symmetric in color and flavor
(pairing a given species with itself), its Dirac structure must be
antisymmetric. We also restrict ourselves to rotationally invariant
pairing, which leaves three possible Dirac structures for the
secondary pairing: $C$, $C\ga_5$, and $C\ga_0\ga_5$.
We give a detailed treatment of
secondary pairing in the $C\ga_0\ga_5$ channel,
because it is the only one that, 
for a single color and single flavor,
is predicted to be attractive under an NJL interaction based on single
gluon exchange (Ref.~\cite{Alford:2002rz}, last 4 lines of Table 2).
We work at zero temperature throughout.

Our results confirm the 
conclusion of Ref.~\cite{Hong} concerning the parametric form of
the enhancement of the
secondary pairing when it operates on quadratically gapless quasiparticles.
We find that for a coupling strength
$G_s$ in the secondary channel, $\De_s\propto G_s^2$ as compared to
the BCS result $\De_s\propto \exp(-C/G_s)$. This was predicted by Hong
\cite{Hong} (note that our $G_s$ should be identified with
Hong's effective interaction strength $\ka$, not with his ``$G_s$'').
This result can also be understood by a simple argument
in the NJL model \cite{Krishna} (see end of section \ref{sec:results}).

\section{Two-species pairing formalism}
\label{sec:secondary}

Our model contains two species of quark with
primary pairing leading to anomalous self-energy
\begin{equation}
\langle \psi_a C\gamma_5 \psi_b \rangle_{1PI} = \Delta_p (\si_{1})_{ab} \ .
\label{primary}
\end{equation}
The secondary pairing that we will discuss in most detail is the
Dirac $C\ga_0\ga_5$ channel,
\begin{equation}
\langle \psi_a C\gamma_0\gamma_5 \psi_b \rangle_{1PI}= \Delta_s \delta_{ab} \ .
\label{secondary}
\end{equation}
The two-dimensional species space, indexed by $a,b=1,2$, corresponds to the
blue-up/red-strange sector of the combined color-flavor space
in full QCD.
As in the gCFL phase, the primary pairing is between species 1 and species 2,
with the form $(1,2) + (2,1)$ given by the Pauli matrix $\si_1$.
It is symmetric in the color-flavor
indices (because it is antisymmetric in both color and flavor) and 
antisymmetric in the Dirac indices. It has spin zero (no spatial indices).
The secondary pairing pairs each species with itself, and is also
antisymmetric in the Dirac indices, and has spin zero.
However it vanishes in the limit of
massless quarks \cite{Alford:2005qw} because it pairs a left-handed quark
with a right-handed quark. (For massless quarks with equal and 
opposite momenta this corresponds to pairing quarks with parallel spins,
which would have to give a spin-1 state.)
This means that $C\ga_0\ga_5$ secondary pairing can only occur if at least one
of the flavors is massive. Fortunately that is the situation
we are interested in, since one of our quarks is light (blue up)
and the other is heavier (red strange).

\new{
One might ask whether we have missed any important physics by
using a model that only represents one sector of the gCFL pairing
pattern. For example, there might be secondary pairing in other
sectors that could feed back into the gap equations in our sector.
However, as we will see below, even within our sector the feedback
of the secondary pairing on the primary pairing is negligible, because
the secondary pairing is so small, so there is no reason to expect
significant contamination from other sectors.
}

To treat the quark-quark condensation, which violates
fermion number and allows quarks to turn into antiquarks,
we use Nambu-Gor'kov spinors, which
incorporate particles and antiparticles into the same spinor,
$(\psi, \bar\psi^T)$. Our fermion fields $\chi$ are
therefore 16-dimensional, arising from a tensor
product of the 4-dimensional Dirac space, the 2-dimensional color-flavor space,
and the Nambu-Gor'kov doubling.

The action for the quarks is
\begin{equation}
A =\frac{1}{2} \int \chi(p)^\dagger S^{-1}(p) \chi(p) 
  \,\frac{d^4p}{(2\pi)^4}\ .
\end{equation}
The inverse propagator $S^{-1}(p)$ is a $16$ by $16$ matrix:
\begin{eqnarray}
&&S^{-1}(p_0,\vec p)=\nonumber\\
&&\hspace{-3em} 
\left(\begin{array}{c@{\hspace{0em}}c}
 (p_{\nu}\gamma_0\gamma^{\nu}\!+\!\bar\mu)\otimes\one
  \!-\!\dmu\otimes\sigma_3
  \!-\!\gamma_0\otimes M
 &\Delta_p C\gamma_5\otimes\sigma_1
  \!+\!\Delta_s C\gamma_0\gamma_5\otimes\one\\
 (\Delta_p C\gamma_5\otimes\sigma_1
  \!+\!\Delta_s C\gamma_0\gamma_5\otimes\one)^T
 &[(p_{\nu}\gamma_0\gamma^{\nu}
  \!-\!\bar\mu)\otimes\one
  \!+\!\dmu\otimes\sigma_3\!+\! \gamma_0\otimes M]^T
\end{array}\right),
\label{fermion_matrix}
\end{eqnarray}
where in each tensor product the first factor lives in the Dirac space,
and the second factor in the color-flavor space. 
The two species have an average chemical potential $\bar\mu$,
and a chemical potential splitting $\dmu$ which corresponds to
the color and electrostatic potentials that enforce neutrality
in the gCFL phase.
The quark mass matrix is
\begin{equation}
M=\left(\begin{array}{cc}
m_u&0\\
0&m_s
\end{array}\right).
\end{equation}
We can obtain the 4 branches (each 4-fold degenerate)
of the full quark dispersion relation,
including the effects of both primary and secondary pairing, by
finding the values of the energy $\eps(\vec p)$ at which there is a
pole in the full propagator
\beq
\det S^{-1}(\eps(\vec p),\vec p) = 0 \ .
\eeq
If we index these solutions by a label $\al=1\ldots 16$ then
the grand canonical potential (or, loosely, the free energy) is
\begin{equation}
\Omega(m_u,m_s,\bar\mu,\dmu,\Delta_p,\Delta_s)=
  - \int^\La \sum_\al 
   \Bigl|\epsilon_\al(\vec p,m_u,m_s,\bar\mu,\dmu,\Delta_p,\Delta_s)\Bigr|
   \frac{d^3p}{(2\pi)^3} 
  + \frac{\Delta_p^2}{G_p}+\frac{\Delta_s^2}{G_s} \ .
\end{equation}
The effective couplings $G_p$ and $G_s$ are determined by details of the
NJL model interaction. In this paper we will simply treat them as
parameters. Our aim is to see how the secondary pairing $\De_s$
depends on the coupling $G_s$ in the secondary channel.
\new{We do not attempt to use ``realistic'' values for
$G_s$ (as might be obtained from commonly used NJL models of QCD)
because we will be able to show that
for any reasonable value of the secondary coupling 
(i.e for $G_s\lesssim\half G_p$)
the secondary gap is too small to generically resolve the
magnetic instability of a gapless phase.}

Our procedure is as follows.
\begin{enumerate}
\item Choose values of $\mubar$, $m_u$, and $m_s$ that are appropriate for
quark matter in the core of a compact star, and an arbitrary ultra-violet 
cutoff $\La$, significantly larger than $\mubar$.
We used $\mubar=500~\MeV$ with
$m_u=0$ to 100~\MeV, $m_s = 160$ to $300~\MeV$, and cutoff
$\La=800$ or $1000~\MeV$.

\item Choose the desired 
gap parameter $\De_p$ for the primary pairing. We varied $\De_p$
from 25~\MeV\ to 75~\MeV. To obtain a given value of $\De_p$ one must
choose a value of the coupling $G_p$
so that the free energy has a minimum at that value
of $\De_p$. As we vary $G_p$ we must also vary $\dmu$
so that quasiparticle dispersion relation after primary pairing is 
always quadratically gapless. To simplify this process we first
do it at $m_u=m_s=0$, where we simply have to set $\dmu=\De_p$ to
obtain quadratically gapless dispersion relations. I.e., we solve
the gap equation
\begin{equation}
\frac{2\Delta_p}{G_p}=\frac{\p}{\p \Delta_p}
 \int^\La \sum_\al 
  \Bigl|\eps_\al(p,m_u=0,m_s=0,\bar\mu,\dmu=\Delta_p,\Delta_p,\Delta_s=0)\Bigr|
 \frac{d^3p}{(2\pi)^3}
\label{gapeq_primary}
\end{equation}
to obtain $G_p$\new{$(m_s=0)$}. We then turn
on the desired value of $m_s$, and retune $\dmu$ to obtain
quadratically gapless dispersion relations. \new{It turns out that
if we now use the primary gap equation (including the non-zero $m_s$)
to determine what value of $G_p$ gives the desired 
primary gap parameter $\De_p$, the result is only slightly different
(by $\lesssim 10\%$, typically) 
from the value of $G_p(m_s=0)$. So in our
results we actually use $G_p(m_s=0)$.}
Note that any ``error'' in $G_p$ is really just a rescaling of
$\be$  in the secondary pairing gap equation \eqn{secondary}, whose
only effect would be to slightly shift the line in Fig.~\ref{fig:Delta_s}.

\item Study how the secondary pairing depends on $G_s$. Recall that
$G_p$ and $G_s$ both come from the same underlying NJL interaction
with some coupling $G$.
They arise from Fierz rearrangements of that interaction, so they
are both of order $G$, with different numerical coefficients.
It is therefore natural to define the ratio of the secondary to the
primary effective coupling strength
\beq
\be = G_s/G_p \ ,
\label{beta}
\eeq
and to study how $\De_s$ depends on $\be$. We expect $\be \lesssim 1$
since the secondary pairing is by definition weaker than the primary pairing.
For the $C\ga_0\ga_5$ secondary channel in a single-gluon-based
NJL interaction, $\be\sim \half$ \cite{Alford:2002rz}.
Our final step is therefore to solve the gap equation for the
secondary pairing strength $\De_s$ as a function of $\be$,
\begin{equation}
\frac{2\Delta_s}{\beta G_p}=\frac{\p}{\p \Delta_s}
 \int^\La \sum_\al 
  \Bigl|\eps_\al(p,m_u,m_s,\bar\mu,\dmu,\Delta_p,\Delta_s)\Bigr|
 \frac{d^3p}{(2\pi)^3}
\label{gapeq_secondary}
\end{equation}
\new{In principle one should solve coupled gap equations
\eqn{gapeq_primary} and \eqn{gapeq_secondary}, but we assume that 
$\De_s\ll\De_p$
so, as we see explicitly below,
the back-reaction of the secondary pairing on the primary gap
is negligible.}

\end{enumerate}

\section{Results for the $C\ga_0\ga_5$ channel}
\label{sec:results}

\begin{figure}[htb]
\begin{center}
\includegraphics[width=0.8\textwidth]{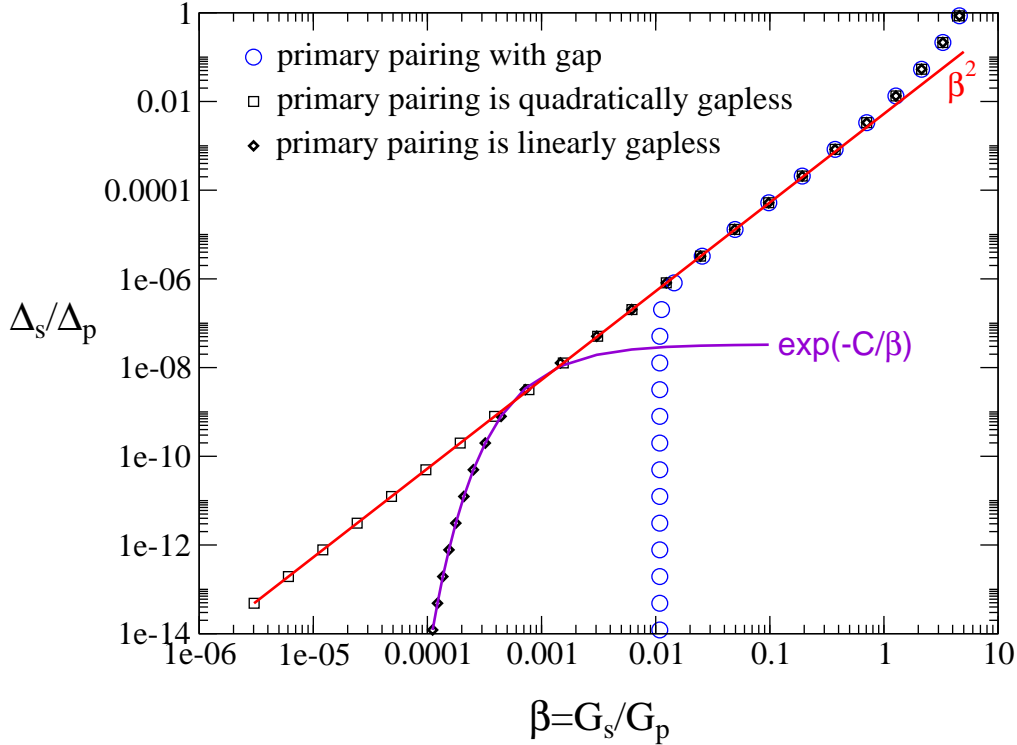}
\end{center}
\caption{\small The ratio of the secondary to primary
pairing gap parameters $\Delta_s/\Delta_p$,
as a function of the ratio of the
coupling strengths in the primary and secondary channels
 $\beta=G_s/G_p$.
In realistic QCD-like interactions we expect $\beta\lesssim\half$,
\new{and our original assumption that $\De_s\ll\De_p$ turns out
to be valid for all $\beta$ in this range.}
For primary pairing with a quadratically gapless dispersion relation
(squares) our results fit $\Delta_s\propto \beta^2$, 
as predicted in Ref.~\cite{Hong}.
For primary pairing with linearly gapless dispersion relation 
(small diamonds) we recover the BCS relation 
$\Delta_s\propto\exp(-C/\beta)$ at low $\beta$.
If the primary pairing leaves a gap (circles) then secondary pairing disappears
at low coupling.
}
\label{fig:Delta_s}
\end{figure}


We present detailed results for the case
$\bar\mu=500~\MeV$, $m_u=0$, $m_s=300~\MeV$, with cutoff $\Lambda=800~\MeV$.
For a primary coupling $G=7.625\ee{-6}~\MeV^{-2}$, which gives
$\Delta_p=75~\MeV$, we tuned $\dmu$ to 
$\dmu_{\rm quad} = -26.54544010627093~\MeV$ to obtain
quasiquark dispersion relations that were quadratically gapless
to within $10^{-13}~\MeV$.
We then calculated
the secondary pairing as a function of $\be$. 
\new{To confirm that the secondary pairing has negligible effect on the
primary gap equation, we took
the strongest secondary channel coupling, $\beta=1$, 
which gave $\De_s=0.555~\MeV$, and
re-solved the primary gap equation, 
including this value
of $\De_s$ in the primary gap equation. This led to a shift
$\de\De_p=0.003~\MeV$, which is certainly negligible
relative to $\Delta_p=75~\MeV$.
As well as the gapless case,
}
we also studied a slightly
less negative value of $\dmu$, for which the quasiparticles had a small
gap, and a slightly more negative value, for which there were
two linearly gapless points.
The results are plotted in Fig.~\ref{fig:Delta_s}.

\begin{figure}[htb]
\begin{center}
\includegraphics[width=0.8\textwidth]{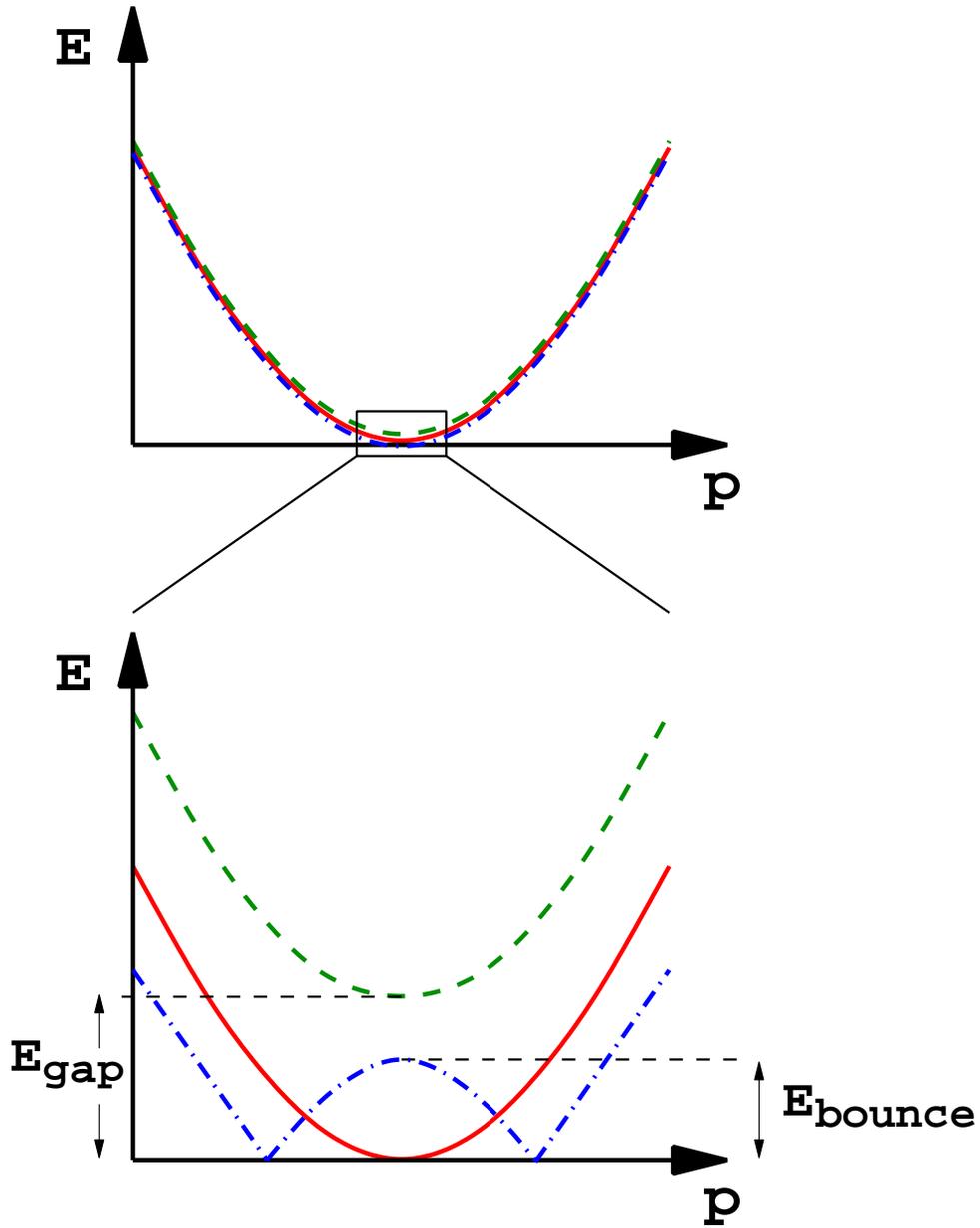}
\end{center}
\caption{\small 
Schematic illustration of the dispersion relations 
of the quasiquarks (before secondary pairing) that underly the three
sets of data plotted in Fig.~\ref{fig:Delta_s}. At higher energy scales,
all three dispersion relations look quadratically gapless (upper panel).
At energies $\lesssim E_{\rm gap}$ it is possible to distinguish
that one case is gapped (dashed line in lower panel).
At energies $\lesssim E_{\rm bounce}$ it is possible to distinguish
that the other has two linearly gapless points 
(dash-dotted line in lower panel).
}
\label{fig:schematic}
\end{figure}

\begin{enumerate}
\item 
\underline{Primary pairing giving exactly quadratically gapless quasiquarks}
  (squares in Fig.~\ref{fig:Delta_s}).\\[1ex]
This occurs when $\dmu=\dmu_{\rm quad}$. Our NJL calculations of
the secondary pairing follow
the expected result, $\De_s/\De_p=A\be^2$ (our fit, straight
solid line, is $A=5.287\ee{-3}$), 
all the way down to the lowest secondary couplings that we could probe.
Note, however, that in the physically relevant range of couplings,
$\be < 1$, the secondary pairing is still suppressed
relative to the primary pairing by a factor of 100 or more.
(On the straight line fit, $\De_s/\De_p(\be\!=\!1) = A =5.287\ee{-3}$).

\item \underline{Primary pairing giving gapped quasiquarks}
  (circles in  Fig.~\ref{fig:Delta_s}).\\[1ex]
We plot the case $\dmu=-26.5454300$, for which the gap in the
spectrum is $E_{\rm gap}\approx 10^{-5}~\MeV$. 
In this case the quasiquark spectrum looks
quadratic at higher energies than this
(see dashed line in schematic plot of dispersion relations,
Fig.~\ref{fig:schematic})
and so we expect the secondary pairing
will only ``notice'' the gap if $\De_s\lesssim 10^{-5}~\MeV$
i.e. $\De_s/\De_p\lesssim 10^{-7}$, which corresponds to secondary coupling
$\be\approx 0.01$.
The results of the explicit NJL
calculation confirm this. For $\be>0.01$ we obtain a quadratic dependence
$\De_s/\De_p\propto\be^2$, 
since the primary quasiquarks look quadratically
gapless. At $\be\approx 0.01$ the secondary pairing falls to zero, since there
are no quasiquark modes within  $\De_s$ of the Fermi surface.

\item \underline{Primary pairing giving linearly gapless quasiquarks}
  (diamonds in  Fig.~\ref{fig:Delta_s}).\\[1ex]
We plot the case $\dmu=-26.5454402$, for which the height of the
``bounce'' in the primary quasiquark dispersion relation 
is $E_{\rm bounce}\approx 10^{-7}~\MeV$. The quasiquark spectrum looks
quadratic at higher energies than this
(see dash-dotted line in schematic plot of dispersion relations,
Fig.~\ref{fig:schematic}),
and so we expect the secondary pairing
will only ``notice'' the linearity if $\De_s\lesssim 10^{-7}~\MeV$, 
i.e. $\De_s/\De_p\lesssim 10^{-9}$, which corresponds to secondary coupling
$\be\approx 0.0003$.
The results of the explicit NJL
calculation confirm this. For $\be>0.0003$ we obtain
$\De_s/\De_p\propto\be^2$, since the primary quasiquarks look quadratically
gapless at energies greater that $\De_s$.
At $\be\approx0.0003$ the secondary pairing changes to the BCS
form, $\Delta_s/\Delta_p = A\exp(-C/\beta)$
(our fit is $A=3.333\ee{-8}$, $C=0.001646$), since the
quasiquark modes within  $\De_s$ of the Fermi surface have linear
dispersion relations, like those of unpaired fermions.
\end{enumerate}

We have checked that the results described above are generic.
We have repeated the calculation for values of $m_s$ from 160 to 300~\MeV,
$m_u$ from 0 to 100~\MeV, and
$\De_p$ from 25 to 75~\MeV, for two different
values of the cutoff $\La=800$ and 1000~\MeV. In every case we found
$\De_s/\De_p \lesssim 0.01$ at $\be=1$.

\clearpage

There is a  a simple calculation in the NJL
formalism, pointed out by K. Rajagopal \cite{Krishna}, which shows
the origin of Hong's predicted $\De_s\propto\be^2$ behavior for quadratically
gapless quasiquarks, and moreover enables us to see exactly how the 
$C\ga_0\ga_5$ gap is suppressed as $m_s\to 0$. We first write
down the generic form of the gap equation,
\beq
1 = G \int^\La \frac{1}{\sqrt{E(p)^2 + \De^2}} p^2 dp \ .
\label{gapeq}
\eeq
In the typical case where the quarks before pairing
have linearly gapless dispersion relations $E(p)\propto |p-\mu|$, 
the integral diverges logarithmically as $\De\to 0$, and
we obtain the BCS result
\beq
1 \approx G\mu^2 \log(\La/\De) \quad\Rightarrow\quad
 \De \approx \La \exp(-1/(G\mu^2)) \ .
\eeq
In the case of a quadratically gapless dispersion relation,
$E(p)\propto (p-\mu)^2/\De_p$, if we  assume
(without justification, at this point) that the secondary
pairing gap equation also takes the form \eqn{gapeq},
with $\De$ replaced by $\De_s$, then the
integral will diverge as an inverse square root of $\De_s$,  
so the gap equation for secondary pairing
would become (using \eqn{beta})
\beq
1 \propto G_s\mu^2\sqrt{\De_p/\De_s} \quad\Rightarrow\quad
\De_s/\De_p \propto \mu^4 G_p^2\,\be^2 \ .
\eeq
which shows the $\De_s\propto\be^2$ behavior.

The full expression for the integrand on the right-hand-side of the
gap equation in the $C\ga_0\ga_5$ channel is very complicated, and we
have not been able to find analytic approximations that
would reduce it to a simple form that could be compared with \eqn{gapeq}.
However, the integrand is dominated by the contribution from the
lowest-energy quasiparticle in the range of momenta where it
would become quadratically gapless at $\De_s=0$, and we have
numerically
evaluated that contribution in the case $m_u=0$,
and found that it is well approximated 
(to within about 10\% for the values of $\mu,m_s,\De_p$
studied in this paper) by
\beq
 \frac{m_s^2}{2\mu^2} \left(\frac{(p-p_F)^4}{\De_p^2}
  + \frac{m_s^2}{\mu^2}\De_s^2 \right)^{-\half}\ .
\label{approx}
\eeq
If we use this expression in the gap equation, we find
\beq
1 \propto G_s\mu^2\left(\frac{\De_p}{\De_s}\right)^{1/2}
   \left(\frac{m_s}{\mu}\right)^{3/2}\quad\Rightarrow\quad
\De_s/\De_p \propto  m_s^3 \mu G_p^2\,\be^2 \ .
\label{secgap}
\eeq
This shows both the $\be^2$ dependence, and the $m_s^3$ suppression
of the $C\ga_0\ga_5$ secondary pairing in the limit $m_s\to 0$.
\new{We have checked the $m_s$ dependence by varying $m_s$
in our numerical calculations, and found that it is well-described by
Eq.~\eqn{secgap}.}

For secondary pairing in the other channels that are not suppressed in
the $m_s\to 0$ limit (such as $C\ga_5$ \cite{Hong}) the factors of
$m_s^2/\mu^2$ in \eqn{approx} would be absent and there would be
a prefactor of $\mu^4$ instead of $m_s^3\mu$ in \eqn{secgap}.
For $C\ga_0\ga_5$
self-pairing of a single isolated flavor, as considered in
Ref.~\cite{Alford:2002rz}), where the unpaired quasiparticles have
linear dispersion relations, the gap integral has a BCS divergence.
In that situation the factors of $m_s^2/\mu^2$ are exponentiated 
to give a much more severe
suppression of the gap as $m_s\to 0$, as seen in Fig.~4 of
Ref.~\cite{Alford:2002rz}.

\section{Other channels}
\label{sec:others}

The detailed results presented above are all for condensation of the form
$\langle \psi_a C\gamma_0\gamma_5 \psi_b \rangle_{1PI}= \Delta_s \delta_{ab}$,
where the single-species pairing has the same sign for
both species, $(1,1) + (2,2)$.
One could also construct secondary pairing of the form
$\langle \psi_a C\gamma_0\gamma_5 \psi_b \rangle_{1PI}=
\Delta_s (\si_3)_{ab}$ where the two species each self-pair with 
opposite sign, $(1,1) - (2,2)$.
In Ref.~\cite{Alford:2002rz} these possibilities were not distinguished
because the different colors and flavors were independent of each
other. In the current context, however, there is primary pairing 
in the background, 
which connects the two species, so that the quasiquarks that undergo
secondary pairing are not species eigenstates, but a momentum-dependent
superposition of $1$ and $2$. This means that the relative sign of the 
secondary $(1,1)$ and $(2,2)$ pairing is physically important.
We have studied the $C\gamma_0\gamma_5\, (\si_3)_{ab}$ channel,
and we find essentially the same behavior as in the 
$C\gamma_0\gamma_5\, \de_{ab}$ channel: $\De_s/\De_p\lesssim 0.01$
for $\be\lesssim\half$ (first line of table \ref{tab:channels}).

We have also analyzed the other
 Dirac-antisymmetric channels that
are rotational scalars, arbitrarily assigning positive (attractive)
interaction strengths $G_s$ to these channels. The results are summarized
in table \ref{tab:channels}. For each channel
we chose the same parameter values as used in Fig.~\ref{fig:Delta_s},
namely $\mubar=500~\MeV$, $m_s=300~\MeV$, $\De_p=75~\MeV$, $m_u=0$,
$\La=800~\MeV$, and we tuned $\dmu$ to obtain a quadratically gapless
quasiquark.
In most channels we obtained
results similar to those presented in Fig.~\ref{fig:Delta_s}:
as the interaction strength drops from 1,
$\De_s/\De_p$ drops as $\be^2$. The only exception is the
$C\de_{ab}$ channel, where we find that there is no secondary pairing
for $\be\leqslant 1$.
However, among the channels studied
we find significant variation in the strength of the secondary
pairing. The potentially physically relevant region is where the 
secondary channel interaction
strength $G_s$ is smaller than the
primary channel interaction strength $G_p$. In table \ref{tab:channels}
we therefore show $\De_s/\De_p$ at $\be=\half$ and 1. We see that
in the $C\ga_5(\si_3)_{ab}$ and $C(\si_3)_{ab}$ channels
the secondary pairing is about ten times weaker than
the primary pairing when $\be=\half$, and that in order for $\De_s$
to be of the same order as $\De_p$, the secondary channel would
have to be as strongly attractive as the primary channel.

\begin{table}[h]
\begin{tabular}{l@{\qquad}cc@{\qquad\qquad}l@{\qquad}cc}
\hline
  \rule[-1.3ex]{0em}{3.7ex} Channel & \multicolumn{2}{c}{$\De_s/\De_p$~~~~~~~~~} 
& Channel & \multicolumn{2}{c}{$\De_s/\De_p$} \\
\rule[-1.3ex]{0em}{3.7ex}  & $\be=\half$ & $\be=1$ & & $\be=\half$ & $\be=1$\\
\hline
\rule[-1.3ex]{0em}{3.7ex} $C\ga_0\ga_5 \,\de_{ab}$ & 0.0015 & 0.0074 
  &  $C\ga_0\ga_5 \,(\si_3)_{ab}$ & 0.0024 & 0.012 \\
\rule[-1.3ex]{0em}{3.7ex} $C\ga_5 \,\de_{ab}$ & 0.00019 & 0.026 
  &  $C\ga_5 \,(\si_3)_{ab}$ & 0.093 & 0.72 \\
\rule[-1.3ex]{0em}{3.7ex} $C \,\de_{ab}$ & $<10^{-11}$ & $<10^{-11}$ 
 &   $C \,(\si_3)_{ab}$ & $0.067$ & $0.53$ \\
\hline
\end{tabular}
\caption{Secondary pairing in various rotationally-invariant
single-species channels.
Parameters were the same as in section \ref{sec:results}.
The secondary channel would have to be as strongly attractive
as the primary channel ($\be=G_s/G_p=1$) for the secondary pairing to
be comparable to the primary pairing. As $\be$ drops below 1,
$\De_s$ rapidly drops to a small fraction of $\De_p$.
}
\label{tab:channels}
\end{table}

\section{Conclusion}
\label{sec:conclusion}

Our results validate the intuition that when the quark
dispersion relations (including primary pairing) are quadratically
gapless, the very large density of states near the Fermi surface
should lead to a parametric enhancement of the secondary pairing,
relative to BCS pairing, in weakly coupled channels. However, it is
clear that even with this enhancement, the secondary pairing is not
typically of comparable magnitude to the primary pairing. 

For the case of the $C\ga_0\ga_5$ channel that is predicted to be
attractive in single-gluon-exchange-based NJL models, the secondary
pairing is at least 100 times weaker than the primary pairing.  The
other rotationally-invariant secondary channels are predicted to be
repulsive ($G_s<0$), but we have studied how they would behave if they
were attractive. We find that for $G_s/G_p\leqslant\half$, they are at
least ten times weaker than the primary pairing. For $G_s/G_p$
approaching 1 the secondary pairing could become important in certain
channels, but that would essentially correspond to assuming that they
were not secondary after all, and the competition between the
``secondary'' and primary channels would have to be taken into
account.  We therefore conclude that in spite of the great enhancement
provided by the large density of states of a quadratically gapless
quasiquark, secondary pairing will not generically resolve the
magnetic instability of the gapless phase.

\new{This result was obtained under the most favorable conditions 
for secondary pairing (quadratic gapless dispersion relation), 
so we expect that it is robust. For example, in full gCFL matter it
might turn out that secondary pairing affects the neutrality condition 
that it is no longer physically correct to tune $\dmu$ in the
blue-strange/red-up sector so
as to obtain a quadratic dispersion relation. However, as is clear
from Fig.~\ref{fig:Delta_s}, this can only further suppress the 
secondary pairing gap.
This means that our conclusion will also apply to the
(blue-down/green-strange) sector of the gCFL pairing pattern,
}
whose quasiquark is linearly gapless with
an energy scale $E_{\rm bounce}\approx \mu_e/2$ which is typically 
tens of MeV. Our results for linearly gapless quasiquarks (diamonds
in Fig.~\ref{fig:Delta_s}) show that weakly attractive secondary channels
could readily open up a tiny (BCS-suppressed) gap at the linearly 
gapless points. However, in order to resolve the Meissner instability
we would need secondary pairing of the same order as the primary
pairing in this sector as well as in the
blue-up/red-strange sector. When $\De_s>E_{\rm bounce}$ the
quasiquarks will behave as if they were quadratically gapless, so our
analysis also applies to secondary pairing in the blue-down/green-strange
sector, and our conclusions apply there as well.

\bc
{\bf Acknowledgements}
\ec
We thank K.~Rajagopal and D.~Hong for useful discussions.
This research was supported by the U.S. Department of Energy
under grant number DE-FG02-91ER40628.

\end{document}